\documentclass[letterpaper,twocolumn,10pt]{article}
\usepackage{usenix}
\usepackage{epsfig}
\usepackage{graphicx}
\usepackage{amsmath,amssymb,xspace}
\usepackage{color}
\usepackage{tabularx}
\usepackage{flushend}
\usepackage{subfig}
\usepackage{url}
\usepackage{cite}

\newcommand{\eg}{e.g.,\xspace}

\newcommand{\etc}{etc.\@\xspace}
\newcommand{\ie}{i.e.,\xspace}

\newcommand{\unit}[1]{\ensuremath{\, \mathrm{#1}}}
\graphicspath{{./figures/}}

\begin{document}

\date{}

\title{On the Security of Carrier Phase-based Ranging}

\author{\rm Hildur \'{O}lafsd\'ottir, Aanjhan Ranganathan, Srdjan Capkun\\
Department of Computer Science\\
ETH Zurich, Switzerland\\
}

\maketitle


\begin{abstract}
Multicarrier phase-based ranging is fast emerging as a cost-optimized solution for a wide variety of proximity-based applications due to its low power requirement, low hardware complexity and compatibility with existing standards such as ZigBee and 6LoWPAN. Given potentially critical nature of the applications in which phase-based ranging can be deployed (\eg access control, asset tracking), it is important to evaluate its security guarantees. Therefore, in this work, we investigate the security of multicarrier phase-based ranging systems and specifically focus on distance decreasing relay attacks that have proven detrimental to the security of proximity-based access control systems (e.g., vehicular passive keyless entry and start systems). We show that phase-based ranging, as well as its implementations, are vulnerable to a variety of distance reduction attacks. We describe different attack realizations and verify their feasibility by simulations and experiments on a commercial ranging system. Specifically, we successfully reduced the estimated range to less than $3\unit{m}$ even though the devices were more than 50 m apart. We discuss possible countermeasures against such attacks and illustrate their limitations, therefore demonstrating that phase-based ranging cannot be fully secured against distance decreasing attacks.
\end{abstract}

\section{Introduction}
The use of proximity and location information is ubiquitous today in a wide range of applications~\cite{HazasFeb04,schiller2004location}. For example, proximity-based access tokens (e.g., contactless smart cards, key fobs) are prevalent today in a number of systems~\cite{RasmussenNov09,GuptaMar06} including public transport ticketing, parking and highway toll fee collection, payment systems, electronic passports, physical access control and personnel tracking. Furthermore, modern automobiles use passive keyless entry systems (PKES) to unlock, lock or start the vehicle. The vehicle automatically identifies and unlocks when the key fob is in proximity, and there is no need for the user to remove the key from his pocket. By eliminating the need for user interaction, PKES-like systems also offer better protection in scenarios, e.g., where the user forgets to lock the car manually. With the advent of modern cyber physical autonomous systems and the internet of things, the need for proximity and location information is only bound to increase.

Numerous ranging techniques~\cite{LiuNov07} that use radio communication signals have been developed in the recent years. Some techniques are based on estimating the change in the physical characteristics of the signal such as amplitude, phase and frequency. For example, ranging systems based on received signal strength (RSS)~\cite{BahlMar00,XiangSep04} rely on the free-space path-loss propagation model to estimate the distance between two entities. Other ranging techniques estimate distance based on the time-of-flight (\eg roundtrip time of flight (RTOF), time-difference-of-arrival (TDOA))~\cite{ubisense10,zebra10} of the radio frequency signal. 

Most of these ranging techniques are inherently insecure. For example, an attacker can fake the signal strength in an RSS-based ranging system. Similarly, in an ultrasonic ranging system, an attacker can gain an advantage by relaying messages over the faster radio-frequency channel~\cite{SedighpourNov05}. Recently, it was shown that the PKES systems used in automobiles are also vulnerable to relay attacks~\cite{FrancillonFeb11}. In a relay attack, the attacker uses two proxy devices to relay the communications between two legitimate entities without requiring any knowledge of the actual data being transmitted; therefore independent of any cryptographic primitives implemented. Researchers were able to unlock the car and drive away even though the legitimate key was several hundred meters away from the car. Similar relay attacks were demonstrated on other radio-frequency based access tokens (NFC phones~\cite{francis2010practical}, Google Wallet~\cite{RolandSep12}), even though the communication range for many such contactless systems is limited to a few centimeters.

Multicarrier phase-based ranging~\cite{bensky2007} is fast emerging as a cost-optimized solution for a wide variety of proximity-based applications. The low hardware complexity and their low power consumption make them suitable for power-constrained wireless sensor system applications. For example, the advent of internet of things has seen an increasing number of \emph{smart and networked} devices being deployed ubiquitously where low power consumption is a key requirement. Today, multicarrier phase-based ranging solutions~\cite{atmel,rfranging,vasisht2016decimeter} that are compliant with prominent standards such as WiFi, ZigBee~\cite{zigbee} and 6LoWPAN~\cite{6LoWPAN} are already being commercialized (\eg warehouse monitoring, child-monitoring). Given the widespread deployment of 802.11 WiFi networks, several indoor localization and ranging schemes~\cite{abrudan2013time,xiong2015tonetrack,vasisht2016decimeter,exel2013carrier} that use the carrier-phase of the radio signals have been proposed. For example, Chronos~\cite{vasisht2016decimeter} leverages the carrier phase information of the 802.11 WiFi signals to implement a centimetre-level localization and ranging system using commodity WiFi cards. The implications of distance modification attacks in scenarios where these systems are deployed in security-critical applications like access control to automobiles, critical infrastructure, and medical devices are significant and have not been investigated so far. 

Therefore, in this work, we investigate the security of carrier phase-based ranging systems and demonstrate their vulnerability to distance modification attacks by exploiting the inherent physical properties of the signal. We focus on attacks which result in a decrease of the measured distance since these have been shown to be most relevant in a majority of security applications. Specifically, we make the following contributions:

\begin{itemize}
	\item We show that phase-based ranging, as well as its implementations, are vulnerable to a variety of distance reduction attacks. To this extent, we describe three different attack realizations with varying degree of attacker complexity and evaluate their effectiveness under various conditions. 
	\item We demonstrate the attack on a commercial multicarrier phase-ranging system and show that it is feasible to reduce the estimated distance significantly. Specifically, through our experiments we successfully reduced the estimated range to less than 3 m even though the devices were more than $50\unit{m}$ apart.   
	\item We discuss possible countermeasures against these distance decreasing relay attacks and illustrate their limitations. We show how implementing countermeasures such as \eg estimating rough time-of-flight, pseudorandom frequency hopping \etc only increases the system complexity without fully securing against distance decreasing attacks. 
\end{itemize}

The rest of the paper is organized as follows. In Section~\ref{sec:background}, we give a brief overview of phase-based distance measurement technique. We describe our distance decreasing relay attacks in Section~\ref{sec:phy-attacks} and present our experimental results in Section~\ref{sec:exp-evaluation}. In Section~\ref{sec:countermeasures}, we discuss possible countermeasures and their effectiveness in preventing the distance decreasing relay attacks. We present related work in Section~\ref{sec:related-work} and finally conclude in Section~\ref{sec:conclusion}.\\

\section{Background}
\label{sec:background}
\subsection{Phase-based Ranging}
\begin{figure}[t]
\centering 
    \includegraphics[width=0.9\columnwidth]{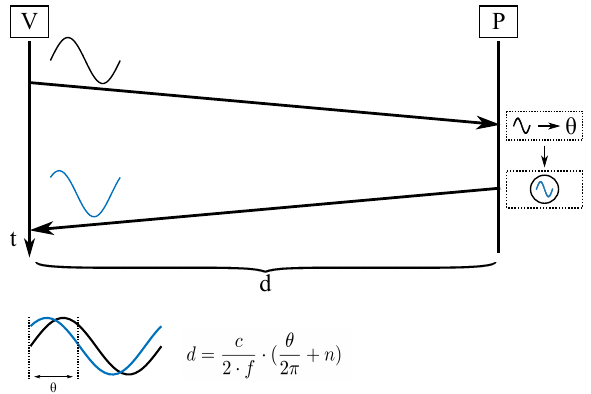} 
    \caption{Phase Ranging: The prover locks its local oscillator to the verfier's signal and transmits it back to the verifier. The verifier then measures the distance based on the difference in the phase of the received signal and its reference signal.}
    \label{fig:phaseranging_protocol}
\end{figure}

In phase-based ranging, two devices A and B measure the distance between them by estimating the phase difference between a received continuous wave signal and a local reference signal. For example, if device A (verifier) is measuring its distance to device B (prover), then the verifier begins ranging by transmitting a continuous wave carrier signal. The prover locks its local oscillator to this incoming signal and transmits it back to the verifier. The verifier measures the distance based on the difference in the phase of the received signal and its reference signal as shown in Figure~\ref{fig:phaseranging_protocol}. If the distance $d$ between the verifier and the prover is less than the signal's wavelength \ie $\dfrac{2\cdot f}{c}$, where $f$ is the frequency of the signal and $c$ is the speed of light, the measured phase difference $\theta$ will be,

\begin{equation}
\theta =4\pi \cdot  \dfrac{d \cdot f}{c}
\end{equation}

In order to unambiguously measure distances greater than the signal's wavelength, it is necessary to keep track of the number of whole cycles elapsed. Therefore, the equation for measuring $d$ becomes,  

\begin{equation}
d =\frac{c}{2 \cdot f }  \cdot (\dfrac{ \theta}{2\pi} +n )
\label{eq:distance_with_ambiguity} 
\end{equation}

where $n$ is an integer which reflects the number of whole cycles elapsed. The need for keeping track of $n$ is eliminated by using continuous wave signals of different frequencies.\\

\subsection{Multicarrier Phase Ranging}
\label{sec:Multicarrier Phase Ranging}

\begin{figure}[t]
\centering
    \includegraphics[width=0.9\columnwidth]{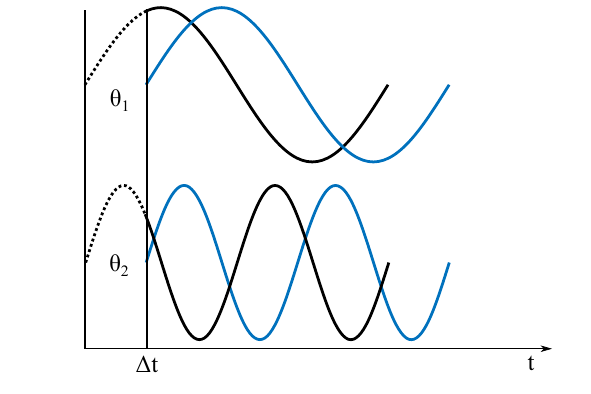} 
    \caption{Multicarrier Phase Ranging: Two signals of different frequencies that travel the same amount of time will experience a different phase shift.}
     \label{fig:mf}
\end{figure}
\begin{figure*}[t]
\centering
\subfloat[The phase of the received signal.]{
    \includegraphics[width=0.45\textwidth]{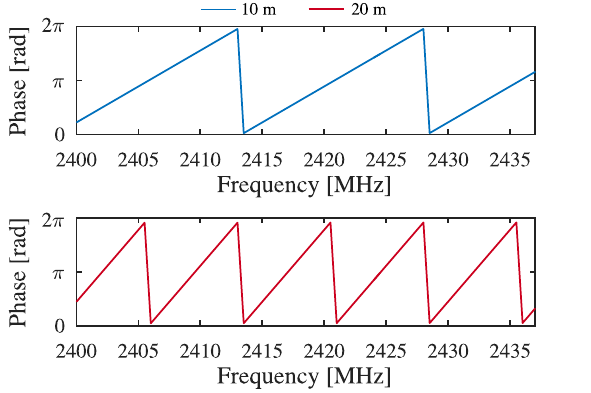} 
   \label{fig:slope-ns}}
   \hspace{0.02\textwidth}
\subfloat[The straightened phase of the received signal.]{
    \includegraphics[width=0.45\textwidth]{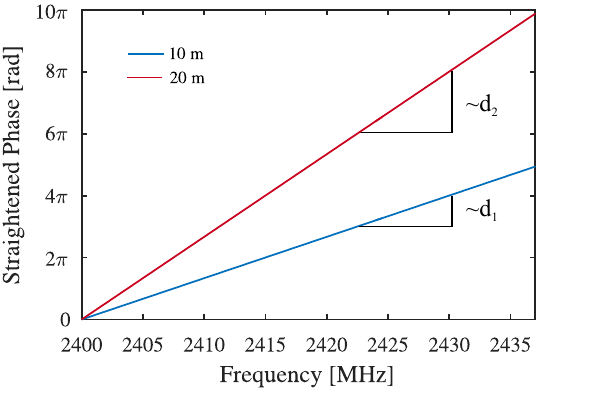} 
    \label{fig:slope-s}}
    \caption{Phase versus frequency if the prover is 10 and 20 m away from the verifier.}
    \label{fig:slope_method}
\end{figure*}

Multicarrier phase ranging systems eliminates the whole cycle ambiguity by transmitting continuous wave signals at different frequencies (Figure~\ref{fig:mf}). For example, the verifier first transmits a signal with a frequency $f_1$ to which the prover locks its local oscillator and retransmits the signal back to the verifier. At the verifier, the measured phase difference between the received signal from the prover and the verifier's own signal for this frequency ($\theta_1$) is given by (from Equation~\ref{eq:distance_with_ambiguity}),

\begin{equation}
	\theta_1 = 2\pi \cdot (\frac{2 \cdot d \cdot f_1}{c} + n)
\label{eq:theta_1}
\end{equation}

The verifier then transmits a continuous wave signal with a frequency $f_2$ and measures the phase difference ($\theta_2$) as previously. 

\begin{equation}
	\theta_2 = 2\pi \cdot (\frac{2 \cdot d \cdot f_2}{c} + n)
\label{eq:theta_2}
\end{equation}

The distance $d$ between the verifier and the prover can be unambiguously measured by combining equations~\ref{eq:theta_1} and~\ref{eq:theta_2}:

\begin{equation}
d = \frac{c}{4\pi} \cdot \frac{\theta_2 - \theta_1}{f_2-f_1}
\label{eq:unambiguous_distance}
\end{equation}

\noindent\textbf{Phase Slope Method:} In real-world, using only two frequencies to measure the phase differences results in poor ranging accuracy. Therefore, it is typical for the verifier to measure the phase differences on more than two frequencies, thereby improving the system's resolution and accuracy. The phase difference measurements ($\theta_i$) for each frequency ($f_i$) can be expressed in the form of:

\begin{equation}
\theta_i = \frac{4\pi}{c} \cdot f_i \cdot d  + N
  \label{eq:phase_slope}
\end{equation}

If the phase differences are plotted on a phase vs frequency curve, the slope of the curve represents the distance $d$ between the verifier and the prover (Figure~\ref{fig:slope_method}). In other words, the above equation can be seen as a straight line with the distance proportional to the slope of the line:

\begin{equation}
  d = \frac{c}{4\pi} \cdot slope
\end{equation}

Figure~\ref{fig:slope-ns} shows the measured phase differences vs frequency for two different distances. The phase-differences are straightened as is shown in Figure~\ref{fig:slope-s} to calculate the effective slope and estimate the distance between the verifier and the prover.\\

\subsection{Commercial Phase Ranging Systems}
Due to their low-complexity and low power requirement, multicarrier phase ranging is  fast emerging as a cost-optimized solution for a wide variety of applications. For example, multicarrier phase ranging has been proposed for the positioning of ultra-high frequency RFID systems~\cite{miesen2012phase,miesen2013360}. More recently, Atmel released a radio transceiver~\cite{atmel} specifically targeting low-power applications and complying with standards such as ZigBee~\cite{zigbee} and 6LoWPAN~\cite{6LoWPAN}. The radio transceiver AT86RF233 is designed for use in industry, scientific and medical (ISM) band applications and implements multicarrier phase-based ranging technique for distance measurement. Further more, leveraging the proliferation of 802.11 WiFi networks and the availability of carrier phase information directly from the network cards~\cite{halperin2011tool}, several indoor localization schemes~\cite{exel2013carrier,abrudan2013time} have been proposed recently. For example, Chronos~\cite{vasisht2016decimeter} leverages the carrier phase information of the 802.11 WiFi signals to implement an indoor localization and ranging system using commodity WiFi cards with centimeter-level precision.


The ranging procedure in these systems is typically divided into control and ranging signals. The control messages are all transmitted using the same preset frequency and is used to set up the necessary parameters and time synchronization for the ranging to take place. In addition, the verifier and prover exchange the results of the ranging using the control channel. The frequencies of the continuous wave signals used in the ranging ranges from $2.324-2.527\unit{GHz}$ with configurable hop sizes of 0.5, 1, 2, 4~MHz.\\

\section{Security of Phase Ranging Systems}
\label{sec:phy-attacks}

In this section, we investigate the security of phase ranging systems with a focus on the physical-layer distance decreasing attacks as these attacks have been shown to be detrimental to a number of security critical applications (\eg NFC payment systems~\cite{RolandSep12,francis2010practical}, keyless entry systems in automobiles~\cite{FrancillonFeb11}).\\

\subsection{Distance Decreasing Relay Attacks}
We consider two devices, a verifier and a prover that are able to communicate over a wireless radio link. The devices implement multicarrier phase measurement for ranging. The verifier measures and verifies the distance claimed by the prover. The verifier is trusted and is assumed to be honest. In this setting, distance decreasing attacks can be mounted in two ways: (i) by a dishonest prover trying to cheat on its distance to the verifier, referred to as an internal attack and (ii) by an external attacker who aims to shorten the distance between the verifier and the honest prover, referred to as a ``distance-decreasing relay attack''.

There are several ways for a dishonest or a malicious prover to mount an internal attack. For example, a malicious prover can cheat on the distance by not locking on to the correct phase when the verifier transmits its interrogating signal (from Figure~\ref{fig:phaseranging_protocol}). The malicious prover can simply respond with a signal that is phase incoherent with the verifier's reference signal; thus resulting in a different distance estimate at the verifier. Such internal attacks can only be prevented by distance bounding~\cite{BrandsMay93} and implementing distance bounding~\cite{rasmussen2010realization,ranganathan2012design,ranganathan2015proximity} require a number of hardware-software modifications that are incompatible with the existing design of phase ranging systems. In this work, we focus on external attackers under the assumption that both the verifier and the prover are trusted and honest. Such a scenario is most applicable to e.g., passive keyless entry systems where the key fob and the car are both trusted and assumed to be honest. However, we note that the presented attacks in this paper can be used by a dishonest prover to decrease its distance to the verifier without any loss of generality. 

Additionally, it is important that the verifier and the prover exchange data that is cryptographically generated. Otherwise, it would be trivial for an unauthorized device to recreate the ranging signals and appear legitimate to the verifier. Throughout this paper, we assume that the verifier and the prover generate and exchange cryptographic data in order to prevent unauthorized ranging attempts.\\

\begin{figure}[t] 
\centering
    \includegraphics[width=0.4\textwidth]{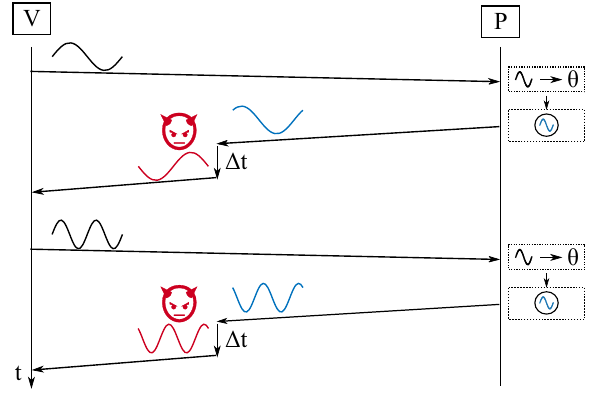} 
    \caption{The verifier's signal travels unaltered from the verifier to the prover. There the prover locks onto the incoming signal and transmits a signal with the same phase back. The attacker intercepts the prover's signal and delays each frequency by the same amount. The verifier calculates an incorrect distance measurement based on the attacker's signal.}
    \label{fig:rollover}
\end{figure}
\subsection{Phase-slope Rollover Attack}
\label{sec:Phase-slope Rollover Attack}
Recall that in a multicarrier phase ranging system, distance $d$ is measured based on the estimated phase differences between two or more carrier frequency signals (Equation~\ref{eq:unambiguous_distance}). Thus, the maximum measurable distance \ie the largest value of distance $d_{max}$ that can be estimated using multicarrier phase-ranging system, depends on the maximum measurable phase difference $\Delta \theta_{max}$ between the two frequency signals. Given that the phase values range from $0$ to $2\pi$, the maximum measurable phase difference between any two frequencies is $\Delta \theta_{max} = 2\pi$. Substituting the values in Equation~\ref{eq:unambiguous_distance} the maximum measurable distance is given by,

\begin{equation}
\begin{split}
  &d_{max} = \frac{c}{4\pi} \cdot \frac{\Delta \theta_{max}}{\Delta f}\\
  &d_{max} = \frac{c}{2} \cdot \frac{1}{\Delta f}
\end{split}
\end{equation}
 
\noindent For example, if the frequency hop size is $2\unit{MHz}$ ($\Delta f$), the maximum distance measurable without any ambiguity is 75~m after which the measured distance rolls over to 0~m. Similarly for frequency hop sizes of 0.5, 1, 2, 4 MHz, the maximum measurable distances are 300, 150, 75 and 37.5 m respectively, beyond which there is a rollover. 

In our phase-slope rollover attack, we demonstrate how an attacker can leverage the maximum measurable distance property of the phase ranging system in order to execute the distance decreasing relay attack.
\begin{figure}[t]
\centering
    \includegraphics[width=.4\textwidth]{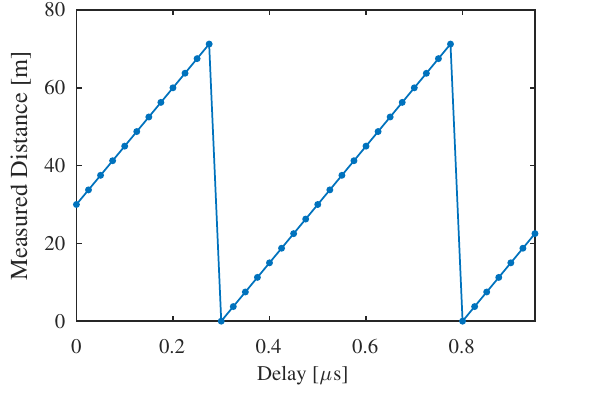} 
    \caption{The verifier and prover are located 30 m from each other and the frequency hop size is 2 MHz (roll over happens at every 500 ns / 75 m). The figure shows the measured distance at the verifier when an attacker uniformly delays all the frequencies by the same amount.}
        \label{fig:phaseslope-rollover}
\end{figure}

The phase-slope rollover attack is illustrated in Figure~\ref{fig:rollover}. The attacker is assumed to be closer to the verifier than the prover. For illustrative simplicity, here we assume that the prover is far away from the verifier or in other words, the verifier and the prover are not in  communication range. During a phase-slope rollover attack, the attacker simply relays (amplify and forward) the verifier's interrogating signal to the prover. The prover determines the phase of the interrogating signal and re-transmits a response signal that is phase-locked with the verifier's interrogating signal. The attacker receives the prover's response signal and forwards it to the verifier, however with a time delay ($\Delta t$). The attacker chooses the time delay such that measured phase differences $\Delta \theta$ between the carrier frequency signals reaches its maximum value of $2\pi$ and rolls over. Considering the previous example of a system with the frequency hop size of $2\unit{MHz}$, the measured phase differences $\Delta \theta$ rolls over every $500\unit{ns}$. Figure \ref{fig:phaseslope-rollover} shows how the measured distance by the verifier changes depending on the delay $\Delta t$ introduced by the attacker. 
In Section~\ref{sec:exp-evaluation}, we demonstrate the feasibility of such an attack on a commercial phase-based ranging system using a experimental setup. Furthermore, we show that an attacker can decrease the estimated distance to the minimum possible distance measurable (depends on sampling rate) by the system irrespective of the true distance of the prover.\\

\begin{figure}[t] 
\centering
    \includegraphics[width=0.4\textwidth]{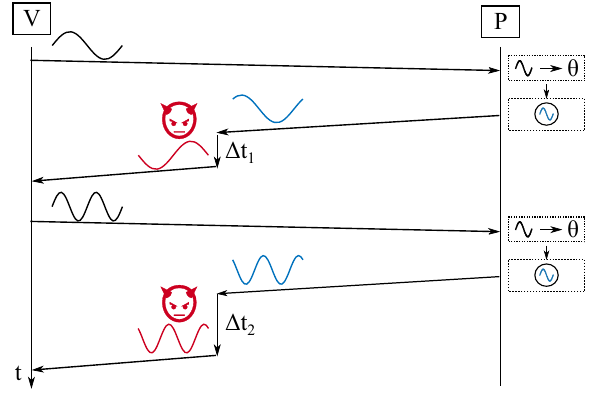} 
    \caption{The verifier's signal travels unaltered from the verifier to the prover. There the prover locks onto the incoming signal and transmits a signal with the same phase back. The attacker intercepts the prover's signal and delays each frequency individually.The verifier calculates an incorrect distance measurement based on the attacker's signal.}
    \label{fig:rfcycleslip}
\end{figure}


\subsection{RF Cycle Slip Attack}

In this section, we describe an alternative way for an attacker to decrease the estimated distance of multicarrier phase ranging systems. In this attack, the attacker manipulates the phase of individual carrier frequencies in order to achieve the required phase difference between the carrier frequencies that will result in a reduced distance estimate at the verifier. This is in contrast to the phase-slope rollover attack described previously, where the attacker simply delays all the carrier frequencies by $\Delta t$ until the effective phase difference between the carrier frequencies exceed the maximum value and rolls over. 

\begin{figure}[t]
\centering
    \includegraphics[width=0.45\textwidth]{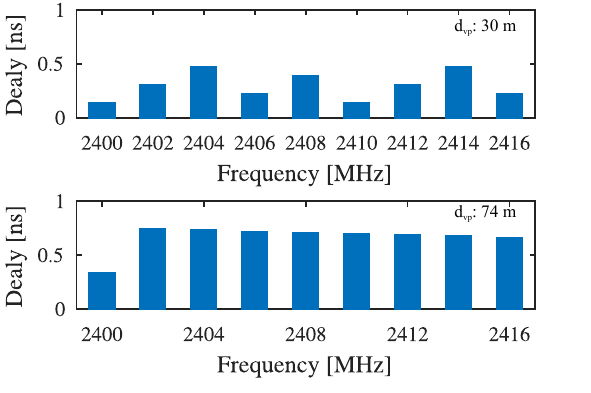} 
    \caption{The delay of each frequency the attacker needs to introduce to decrease the distance to 1 m from  30 and 74 m respectively. Here $d_{vp}$ is prover-verifier distance.}
    \label{fig:results_rfcycleslip}
\end{figure}

In a RF cycle slip attack, the attacker delays each carrier frequency $f_i$ by $\Delta t_i$. Recall that at the verifier, phase difference $\theta_i$ is measured between the prover's response signal and the verifier's reference signal for frequency $f_i$. Thus, an attacker can alter $\theta_i$ by delaying individual carrier signals by an amount that causes each phase measurement to change to a value $\theta_{i}^\prime$. The attacker chooses the new phase, $\theta_{i}^\prime$, for each frequency such that the slope of the phase vs frequency graph decreases and thus decreasing the measured distance. Figure~\ref{fig:results_rfcycleslip} illustrates the delays needed for individual carrier frequencies to cause a particular distance estimate by the verifier. One of the drawbacks of this method is that the attacker needs very high sampling rate. Alternatively, the attacker can use analog delay lines~\cite{SpringerApr01,moon2000all} to realize such a relay attack hardware.\\

\subsection{On-the-fly Phase Manipulation Attack}
\label{sec:On-the-fly Phase Manipulation Attack}
In this section, we present a \textit{real-time phase manipulation} attack, in which the attacker is not required to delay the prover's response signal. In this attack, the attacker manipulates the phase of the prover's response signal by mixing it with specially crafted signal which results in an appropriate phase difference at the verifier. It is important to note that the real-time phase manipulation attacks keeps any possible data exchanged intact independent of the modulation scheme used.

\begin{figure}[t]
\centering 
    \includegraphics[width=.4\textwidth]{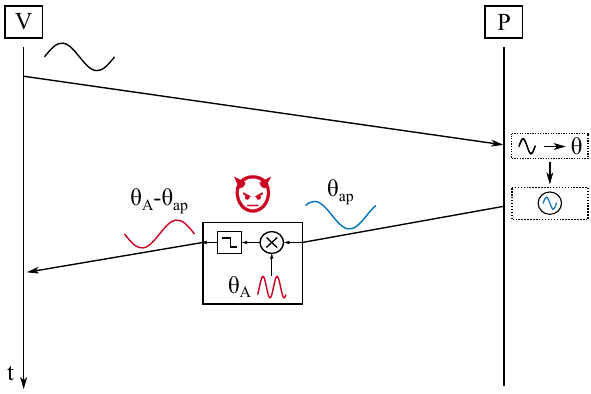} 
    \caption{Attacker phase shifts the prover's signal by first mixing it with a signal of twice the frequency and then low-pass filters the result before transmitting it to the verifier.}
        \label{fig:otf_attack}
\end{figure}

Figure~\ref{fig:otf_attack} illustrates the real-time phase manipulation attack. 
The prover receives the interrogating signal and re-transmits a phase-locked response signal back to the verifier. The prover's response $s_P(t)$ can be expressed as 

\begin{equation}
s_{P}(t) = \cos(2\pi  f  t+\theta_{ap})
\end{equation}

where $f$ is the signal frequency and $\theta_{ap}$ is the received phase of the prover's signal at the attacker. The attacker receives the prover's signal $s_{P}(t)$ and mixes it with a specially crafted signal $s_{if}(t)=\cos(4\pi  f  t+\theta_A)$ before relaying the signal to the verifier. Note that the crafted signal has twice the frequency of the prover's response signal. This is to account for the frequency conversion that occurs during mixing of two signals. The attacker's signal $s_A(t)$ (after filtering high frequency components) that is finally relayed to the verifier can be derived as follows:

\begin{equation}
\begin{split}
s_A(t) &= s_{if}(t)\otimes s_P(t)\\
&=\cos(2\pi ft+\theta_{ap}) \otimes \cos(4\pi ft+\theta_{A})\\
&\stackrel{LP}{=}\frac{1}{2} \cos \Big( 2\pi f t +\theta_A - \theta_{ap}  \Big)
\end{split}
\label{eq:mixing}
\end{equation}
From Equation~\ref{eq:mixing}, we observe that the relayed signal $s_A(t)$ is identical to the prover's response signal except that it is shifted in phase. Recall that (Equation~\ref{eq:unambiguous_distance}), in a multicarrier phase ranging system, the measured distance depends on the change in phase difference measurements between each carrier frequency. Thus, in order to modify the measured distance, the attacker needs to manipulate the phase of each carrier frequency such that it results in a reduced distance estimate. In other words, the attacker has to choose $\theta_A$ such that $\theta_A - \theta_{ap}$ results in a phase difference estimate that corresponds to the reduced distance. 
In order to configure $\theta_A$, the attacker must have apriori knowledge of the phase of the prover's signal when received at the attacker's location ($\theta_{ap}$). The attacker can detect the phase of the verifier's signal when it received it and if the attacker  knows the distance between the attacker and the prover, the attacker can estimate $\theta_{ap}$. An alternative method for the attacker would be to actually detect (\eg using a phase-locked loop) the phase of the prover's response signal. However, this would introduce unnecessary delays\footnote{due to the settling time of PLLs} in the relaying hardware thus making it less favourable for the attacker.\\

\section{Experimental Evaluation}
\label{sec:exp-evaluation}

In this section, we evaluate the feasibility of the above described distance decreasing relay attacks using both commercial phase-ranging systems and simulations. First, we demonstrate the distance decrease relay attack on the commercially available Atmel AT86RF233 radio tra\-nsceiver~\cite{atmel,rapinski2015zigbee} that implements multicarrier phase-based ranging technique. Furthermore, we evaluate the feasibility of the attacks in different environmental conditions (\eg noise, communication range) using simulations.\\

\begin{table}[h]
\centering
\caption{Atmel hardware configuration for the attack}
\begin{tabular}{l|c}
  \hline
  \textbf{Parameter}&\textbf{Value}  \\
  \hline
  Frequency Hop $\Delta f$& $2\unit{MHz}$ \\
  Ranging Frequency Range  & $2.403-2.443\unit{GHz}$\\
  Control Message Frequency & $2.4\unit{GHz}$\\
  No. of Frequencies & $20$\\
  Signal Strength & $-17\unit{dBm}$\\
  \hline
\end{tabular}
\label{tab:atmel-parameters}
\end{table}
\begin{figure}[t]
\centering
  \includegraphics[width=\columnwidth]{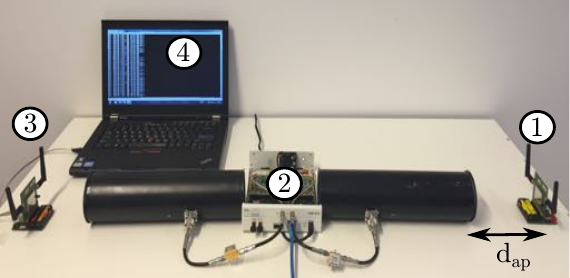}
  \caption{Our experimental setup consists of two multicarrier phase ranging devices  based on the Atmel AT86RF233 radio transceiver, that function as the prover (1) and verifier (3). The distance measured by the verifier is recorded to the connected laptop (4). The attacker's hardware (2) consists of an USRP~\cite{USRPN210} and two directional antennas, one each for receiving the prover's response signal and the other for transmitting the attacker's signal to the verifier. }
  \label{fig:exp-setup-atmel}
\end{figure}

\begin{figure}[t]
    \centering 
    \includegraphics[width=0.9\columnwidth]{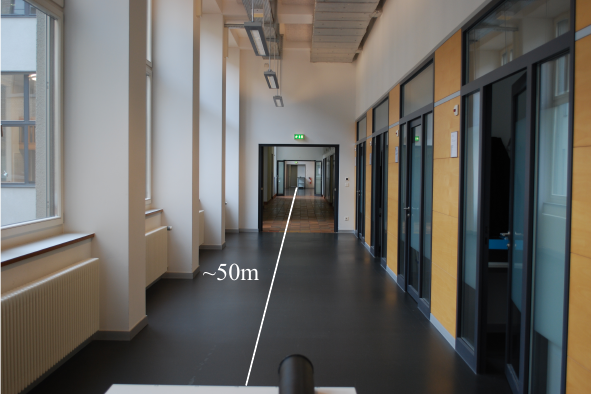} 
    \caption{Hallway in which the experiment took place.}
        \label{fig:hallway}
\end{figure}

\subsection{Practical Demonstration of the Attack}  
Figure~\ref{fig:exp-setup-atmel} shows the experimental setup used in evaluating the feasibility of executing the distance decreasing relay attack on the Atmel phase-ranging system. Our setup consists of two multicarrier phase ranging devices based on Atmel AT86RF233 radio transceiver. One device (1) acts as the prover while the other device (3) takes the role of the verifier. The verifier continuously measures the distance between the prover and itself and outputs the result to the connected laptop (4). The laptop was configured to continuously log the distance measurements. We used Atmel's default setup for configuring the ranging parameters\footnote{For the 50 m attack the transmit power of the Atmel devices was increased to $-10\unit{dBm}$} and list them in Table~\ref{tab:atmel-parameters}.\\ 
\begin{figure}[t]
    \centering 
    \includegraphics[width=0.9\columnwidth]{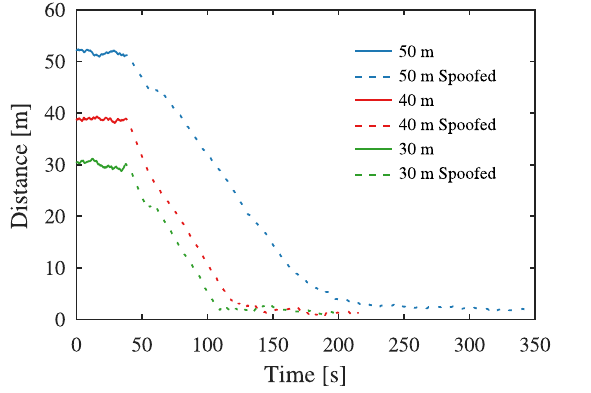} 
    \caption{Effectiveness of the distance decreasing relay attack where the prover and verifier are located 30, 40 and 50 m apart and the attacker attempts to decrease the distance.}
        \label{fig:distance_experiment_30m_40m_50m}
\end{figure}

\subsubsection{Attacker Hardware}
The attacker's hardware (2) consists of an USRP~\cite{USRPN210} and two directional antennas, one used for receiving the prover's response signal and the other for transmitting the attacker's signal to the verifier. The attacker setup was placed close to the verifier while the prover was placed at different distances to the verifier. We implemented the phase-slope rollover attack described in Section~\ref{sec:Phase-slope Rollover Attack} in which the attacker delays all the carrier frequencies until the effective phase difference between the frequencies exceed the maximum value of $2\pi$ and rolls over. The verifier's interrogating signal was left unmodified and the attacker manipulated (delayed and amplified) only the prover's response signal.
In order to minimize the processing delay due to the attacker's hardware, all processing was done directly on the USRP's FPGA, that included receiving, delaying and transmitting the signal. In other words, the host computer of the USRP was bypassed completely and the signal processing was done solely in the FPGA of the USRP. The delay from receiving to transmitting, caused by the USRP hardware, was $536.22\unit{ns}$ with a standard deviation of $1.83\unit{ns}$.
The USRP's host computer was only used to trigger the relay attack and for specifying the amount of delay to introduce into the prover's response signal. The delay was made configurable from the host and tuned at runtime to achieve the desired attack objective.\\

\subsubsection{Experimental Results}
We placed the prover at distances $30\unit{m}$, $40\unit{m}$ and $50\unit{m}$ away from the verifier in an empty hallway. The prover and verifier were in communication range during the experiment and thus were able to estimate their true distance in the absence of the attacker. The results of our experiment are shown in Figure~\ref{fig:distance_experiment_30m_40m_50m}. As can be observed, without the presence of the attacker (solid line), the verifier and the prover estimate their true distance. When the attack is triggered, the verifier's estimated distance begins to reduce. The gradual reduction is due to the verifier averaging the range estimates over a number of samples. We note that the experiment was carried out in a corridor (see Figure~\ref{fig:hallway})with significant interference from other ISM band systems (\eg WiFi). Even in such conditions, our attacker was able to reduce the distance estimate by more than $50\unit{m}$.\\

\subsubsection{Rollover Using Only Amplification}
If two phase-ranging devices are further away from each other than the maximum unambiguous distance that they can measure an attacker can cause a roll-over by simply amplifying their signals. We simulated such and attack on the Atmel AT86RF233 radio transceivers. We placed the devices  at roughly $53\unit{m}$ apart. When the devices were  configured to use a frequency hop size of $2\unit{MHz}$ they correctly estimated their position. However, when configured to use  a  hop size of $4\unit{MHz}$ they incorrectly measured a distance of $15-16\unit{m}$, which is consistent with the rollover being $37.5\unit{m}$. Such an attack is simple to implement but of course the attacker can only reduce the distance rather than spoof the devices to a particular distance  since the measured distance will be determined by the devices actual distance.

\begin{figure}[t]
\centering 
    \includegraphics[width=.4\textwidth]{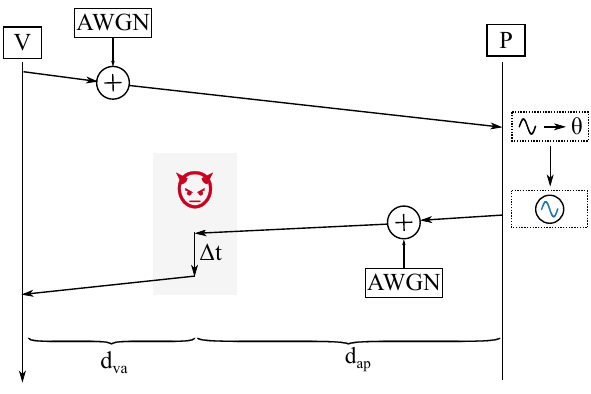} 
    \caption{Simulation of the Phase-slope rollover and  RF cycle slip attack where $d_{va}$ and $d_{ap}$ is the verifier-attacker distance and attacker-prover distance respectively. Additive White Gaussian noise is added to the  verifier's and prover's signal. The attacker delays the prover's signal. For the Phase-slope rollover attack all frequencies are delayed equally but for the RF cycle slip attack each carrier frequency is uniquely delayed.}
        \label{fig:rollover_simulation}
\end{figure}
\begin{figure}[t]
\centering 
    \includegraphics[width=.4\textwidth]{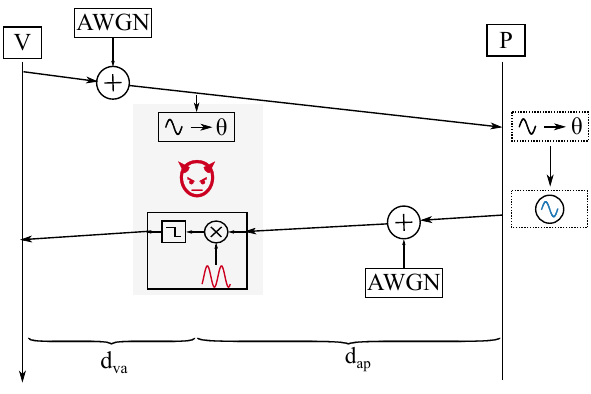} 
    \caption{Simulation of the on-the-fly attack where $d_{va}$ and $d_{ap}$ is the verifier-attacker distance and attacker-prover distance respectively. Additive White Gaussian noise is added to the  verifier's and prover's signal. The attacker estimates the phase of the verifier's signal and uses it and knowledge of the distance to the prover to estimate the phase of the prover's signal when it arrives at the attacker. The attacker then mixes and low-pass filters the prover's signal to achieve the desired phase shift.}
        \label{fig:otf_simulation}
\end{figure}

\subsection{Theoretical Evaluation}
In this section, we evaluate the effectiveness of the distance decreasing relay attack under various channel conditions using simulations.\\

\subsubsection{Simulation Setup}
For the simulations, we implemented the verifier, the prover and the attacker in Matlab. The multicarrier phase-ranging system was modelled exactly as described in Section~\ref{sec:Multicarrier Phase Ranging}. Similar to real-world phase ranging systems, the verifier uses multiple carrier frequencies in the ISM band as the interrogating signal. The range of frequencies used were $2.40-2.48\unit{GHz}$
with a configurable frequency hop of $1\unit{MHz}$ or $2\unit{MHz}$. The phase of the verifier's interrogating signal is selected randomly for each frequency hop to simulate real-world behaviour. The prover measures the phase of the verifier's signal as in a real system and generates its response signal that is phase synced to the verifier's interrogating signal. For evaluating the effectiveness of the attack under noisy channel conditions, white Gaussian noise is added to both the verifier's and the prover's signal. The distances between the verifier, prover and the attacker were simulated by introducing propagation delays in the signal. For example, in order to simulate  a verifier-prover distance of $30\unit{m}$, the signals were temporally shifted by $100\unit{ns}$ before they were processed by the verifier or the prover. 
\begin{figure}[t]
\centering 
    \includegraphics[width=.9\columnwidth]{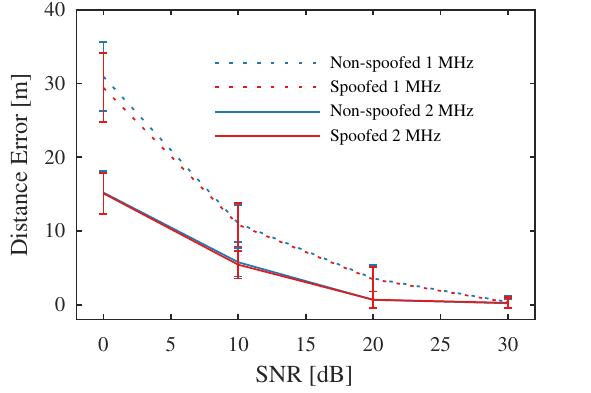} 
\caption{The error in the measured distance in a non-adversarial and when a RF cycle slip attack is performed.  The non-adversarial measurements are when  prover is located 1 m away from the verifier. In the adversarial setting the prover is located 30 m away from the verifier and the attacker tries to reduce this distance to 1 m. In the adversarial setting the prover and verifier are not in communications range}
  \label{fig:RF_cycle_slip_SNR}
\end{figure}
\begin{figure}[t]
\centering 
    \includegraphics[width=.9\columnwidth]{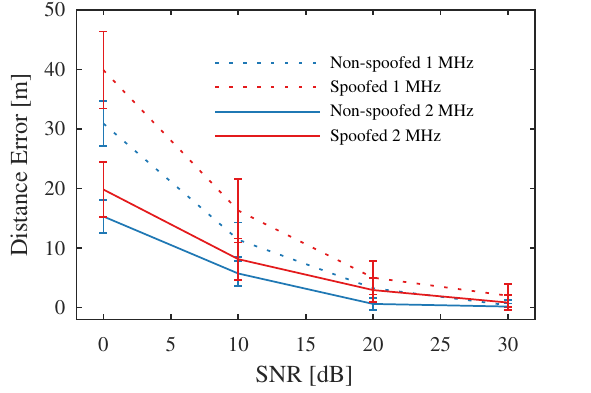} 
    \caption{The error in the measured distance in a non-adversarial and when a OTF attack is performed.  The non-adversarial measurements are when  prover is located 1 m away from the verifier. In the adversarial setting the prover is located 74 m away from the verifier and the attacker tries to reduce this distance to 1 m. In the adversarial setting the prover and verifier are not in communications range}
        \label{fig:OTF_SNR}
\end{figure}
\begin{figure}[t] 
\centering
    \includegraphics[width=.9\columnwidth]{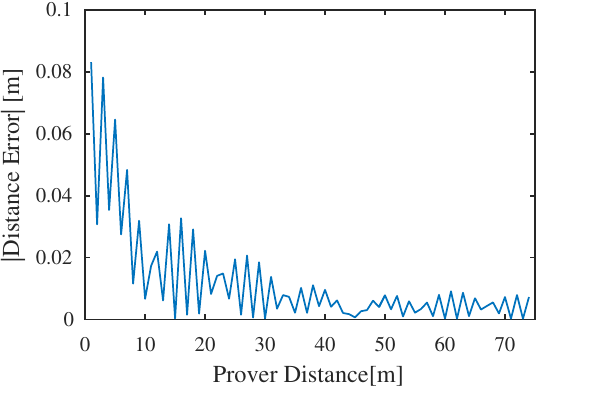} 
    \caption{The effect on measured distance when the verifier and prover are in communications range and  the attacker does not correct for the effect from the prover's signal. The attacker is located 1 m away from the verifier and tries to decrease the distance to 1 m for all prover-verifier distances.}
        \label{fig:interference_prover}
\end{figure}

The attacker was modelled depending on the type of attack evaluated. In the scenario of the phase-slope rollover and the RF cycle slip attack (Figure~\ref{fig:rollover_simulation}), the attacker only received and delayed the response signal from the prover appropriately before relaying it to the verifier. In the case of on-the-fly phase manipulation attack (Figure~\ref{fig:otf_simulation}), the attacker estimates the phase of the verifier's signal to be able to estimate the phase of the prover's signal when it reaches the attacker. The attacker then mixes the received response signal with his locally generated signal as described in Section~\ref{sec:On-the-fly Phase Manipulation Attack} to generate a attack signal that is appropriately shifted in phase in order to reduce the distance estimate while preserving the carrier frequency. The attack signal is low-pass filtered and relayed to the verifier.\\


\begin{figure*}[t]
\centering
\subfloat[The received phase at the verifier if an attacker changes the phase of each carrier randomly.]{
    \includegraphics[width=.9\columnwidth]{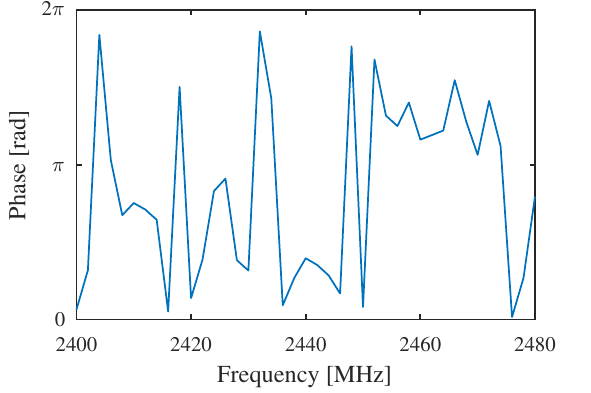} 
   }
   \hspace{0.02\textwidth}
   \subfloat[The verifier attempt to straighten the  phase and linearly fit it.]{
    \includegraphics[width=.9\columnwidth]{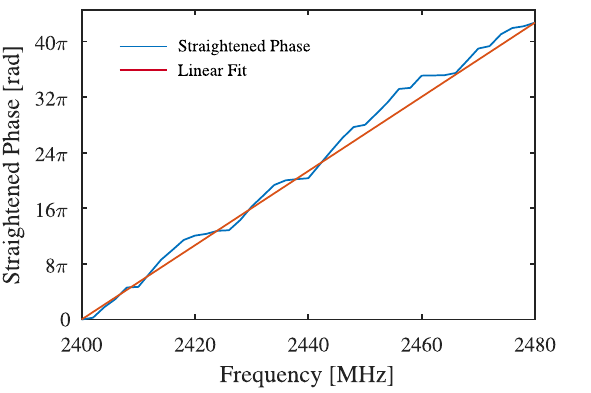} 
   }
   \hspace{0.02\textwidth}
    \caption{The distance estimated based on a random phase will be approximately ~$\frac{d_{max}}{2}$ where $d_{max}$ is the maximum measurable distance.}
        \label{fig:random}
\end{figure*}

\subsubsection{Effect of Channel Noise}
We evaluated the effectiveness of the various distance decreasing attacks described in Section~\ref{sec:phy-attacks}
 under different noise conditions. The evaluations were averaged over 100 different iterations for each SNR value in the set $[0-30]\unit{dB}$. We compared the error in the estimated distance in an adversarial and non-adversarial scenario. The non-adversarial setting was simulated with the prover located $1\unit{m}$ away from the verifier without any attacker present. In the adversarial scenario, an attacker located $1\unit{m}$ away from the verifier relayed the signals between the verifier and the prover. The verifier and the prover were assumed to be out of communication range. Additive white Gaussian noise was added to both the verifier's interrogating signal and the prover's response signal. Figure~\ref{fig:RF_cycle_slip_SNR} and Figure~\ref{fig:OTF_SNR} shows the results for the RF cycle slip attacker and On-the-fly phase manipulation attacker respectively. We simulated the attacks for the commonly used frequency hop size of $1\unit{MHz}$ and $2\unit{MHz}$. As seen in Figure~\ref{fig:RF_cycle_slip_SNR}  for the RF cycle slip attack, there is little difference in the distance error between the adversarial and non-adversarial setting. However, the on-the-fly phase manipulation attacker performs slightly worse than the non-adversarial setting. This is because the attacker must estimate the verifier's phase under noisy conditions and any error in this estimation results in an incorrect phase shift.\\

\subsubsection{Effect of Interference from the Prover}
In certain scenarios, it is common that the verifier and the prover are in communication range and the verifier  also receives the legitimate response signals in addition to the attacker's signals. In this set of experiments, we evaluated the effect of interference caused by the legitimate prover signals on the ability of the attacker to reduce the estimated distance. The amplitude and phase of the received signal at the verifier will depend on both the amplitude and phase of the attacker and the prover signals. For example, if the prover's signal is weaker than the attacker's, the effect on the estimated distance due to the legitimate prover's signal will be minimal. Figure~\ref{fig:interference_prover} shows the deviation in the distance calculated by the verifier for different verifier-prover distances. In our simulations, the attacker was located $1\unit{m}$ away from the verifier and the prover's distance from the verifier was varied. The attacker's objective was always to force the estimated distance to be $1\unit{m}$. It can be seen that the effect is negligible even if the prover is located at a distance of $10\unit{m}$ from the verifier.\\

\subsubsection{Random Phase Manipulation Attack}
An attacker can simply introduce a random phase change to the prover's signal, by either randomly delaying the phase of  individual carrier frequencies or introducing a random phase change in the on-the-fly attack. Figure~\ref{fig:random} shows the received phase at the verifier if such an attack is performed.  A naive phase-ranging system might simply try to linearly fit a slope to the measured phase which will result in an incorrect distance. Depending on the true distance of the prover and verifier, the attacker might thus achieve a distance reduction by simply randomly manipulating the phase. However, a verifier should be able to detect that the received phase is abnormal and thus surmise that any distance calculated from it would be incorrect.\\

\section{Effectiveness of Possible Countermeasures}
\label{sec:countermeasures}

In this section, we discuss possible countermeasures and their effectiveness in preventing the distance decrease attacks described previously.\\ 

\subsection{Frequency Hopping}
In order to execute the distance decreasing attack, the attacker must know the correct carrier frequency or be capable of re-transmitting the entire set of frequencies used for ranging. So, an obvious countermeasure would be to implement pseudo-random frequency hopping. In other words, the verifier and the prover change carrier frequencies based on a shared secret during the ranging process. However, it would be ineffective against attackers capable of listening and transmitting over the entire range of frequencies used by the system. With the widespread availability of low-cost, high-bandwidth amplifiers~\cite{datasheetsminicircuits}, it is reasonable to assume that the attacker would be capable of executing these attacks over the entire range of frequencies used by the multicarrier phase ranging system. Moreover, the attacker can listen to the verifier's interrogating signal that is necessary for the prover to lock and retransmit its response, thereby easily detecting the next frequency used by the verifier and prover to execute the ranging. Thus, a large bandwidth or a pseudo-random frequency hop sequence would be ineffective in preventing distance decreasing attacks.\\

\subsection{Rough Time-of-Flight Estimation}
An alternative countermeasure would be to realize a rough time-of-flight estimation. The verifier and the prover can implement a challenge-response mechanism \ie the verifier modulates challenges in the interrogating signal that is transmitted to the prover. The prover demodulates the challenge, computes a corresponding response and modulates them back on the phase-locked response signal that the prover transmits back to the verifier. Assuming that the signals travel at the speed of light and knowing the prover's processing time, the verifier can estimate a coarse distance by measuring the time elapsed between transmitting the challenges and receiving the responses. It is well established that the precision of the time estimate depends on the system bandwidth~\cite{bensky2007}. Commercially available phase-ranging radio transceivers today are capable of exchanging data at a maximum rate of $2\unit{Mbps}$. Assuming that the transceivers can estimate time-of-flight at this data rate, the maximum achievable precision is $500\unit{ns}$, which translates to a  distance estimate of $150\unit{m}$. This means that, the system would potentially detect attacks in scenarios where the prover is greater than $150\unit{m}$ away from the verifier. 

It is important to note that the time-of-flight estimate would only guarantee whether the prover is within, for example $150\unit{m}$. This still leaves a lot of room for an attacker to execute a distance decreasing attack as phase-ranging would still be required in addition to rough time-of-flight for precise distance estimates. For example, the attacker can still reduce the estimated distance to $1\unit{m}$ even in scenarios where the prover is located $100\unit{m}$ away from the verifier. In order to improve the precision of the time-of-flight estimate, it is necessary to increase the system bandwidth. 
Given that one of the main advantages of multicarrier phase ranging is its low-complexity and cost, increasing the bandwidth for better time-of-flight estimate will potentially make the phase-ranging system redundant.\\ 
\begin{figure}[t] 
\centering
    \includegraphics[width=.4\textwidth]{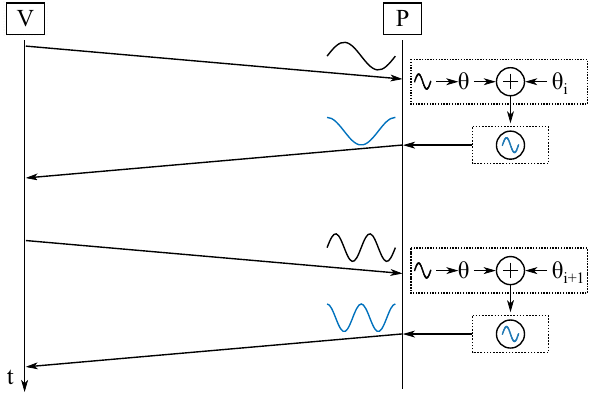} 
    \caption{The prover adds a phase offset to the received phase, which is known to the verifier.}
        \label{fig:attack2_sines}
\end{figure}

\subsection{Phase-shifted Response Signal}
Even though implementing a time-of-flight estimation prevents rollover attacks, it is ineffective against an attacker capable of on-the-fly phase manipulation. As described previously, in an on-the-fly phase manipulation attack, the attacker does not reduce the estimated distance by delaying the signals. The attacker mixes a locally generated intermediate frequency signal with the response signal in real-time to generate the attack signal. In order to generate the intermediate frequency signal, the attacker must know the phase of the incoming response signal. The attacker can estimate the phase of the incoming response signal based on the prover's distance. The prover can potentially leverage this requirement for the attacker and introduce additional phase-shifts in its response signals. The phase-shifts introduced by the prover can be agreed apriori with the verifier and can be accounted for during the distance estimation. The attacker cannot guess the phase-shift that the prover introduces and thereby cannot generate a corresponding mixing signal to execute the distance reduction attack and thereby will result in large fluctuations in the measured phase difference across the carrier frequencies. Recall that, in an non-adversarial setting, the phase difference between the carrier frequencies would vary linearly.

However, an attacker can always detect the phase of the response signal and accordingly generate the mixing signal. Due to the required precision of the phase estimates, the verifier and the prover transmit their interrogating and response signals for a long duration of time $60-100\unit{\mu s}$, in order to allow the phase-locked loop to converge to a precise value. This gives significant time for the attacker to detect the phase of the response signal and generate the necessary mixing signal for the distance decreasing attack. Furthermore, it is important to note that, this technique does not prevent the rollover attacks and hence has to be combined with rough time-of-flight estimation technique. This further increases the complexity of the system, thereby making other ranging techniques such as UWB-IR better suited for security critical applications.\\

\section{Related Work}
\label{sec:related-work}
In this section, we discuss relevant related work in physical-layer security of wireless ranging systems beginning with the works closest to ours. Physical-layer attacks exploits the physical properties of the radio communication system and are therefore independent of any higher layer cryptographic protocols implemented. Several attacks ranging from simply relaying the signal between two legitimate nodes to injecting messages at the physical layer were demonstrated in the past. Clulow et al.~\cite{clulow2006so} introduced physical-layer attacks such as early detect and late commit attacks. In an early detect attack, the attacker predicts the data bit before receiving the entire symbol while in a late commit attack, the attacker leverages the ability of the receiver to decode the bit even though the entire symbol has not been correctly received. The feasibility of these attacks on a ISO 14443 RFID was demonstrated in~\cite{HanckeApr08}.

For short and medium-distance precision ranging and localisation, ultra-wide band (UWB) and chirp spread spectrum (CSS) emerged as the most prominent techniques~\cite{SahinogluDec06} and were standardized in IEEE 802.15.4a~\cite{IEEE802154a07} and ISO/IEC 24730-5~\cite{iso14443}. Flury et al.~\cite{FluryMar10} evaluated the security of impulse radio ultra wide-band PHY layer. The authors demonstrated an effective distance decrease of 140 m for the mandatory modes of the standard. Poturalski et al.~\cite{PoturalskiApr11,PoturalskiSep10} introduced the Cicada attack on the impulse radio ultra wide-band PHY. In this attack, a malicious transmitter continuously transmits a ``1'' impulse with power greater than that of an honest transmitter. This degrades the performance of energy detection based receivers resulting in distance reduction and possibly denial of service. Ranganathan et al.~\cite{RanganathanApr12} investigated the security of CSS-based ranging systems and demonstrated that an attacker would be able to effectively reduce the distance estimated by more than 600~m.

To the best of our knowledge, the security of phase ranging systems have not been evaluated in literature. However, there have been several works~\cite{miesen2012phase,farnsworth2001high,salido2013multipath,zhang2014unambiguous,rapinski2015zigbee} that evaluated novel high precision distance measurement techniques using carrier phase of a signal.\\

\section{Conclusion}
\label{sec:conclusion}

In this work, we investigated the security of multicarrier phase-based ranging systems and demonstrated its vulnerability to distance decreasing relay attacks. We demonstrated both through simulations and real world experiments that phase-based ranging is vulnerable to a variety of distance reduction attacks. We showed that an attacker can reduce the distance measured by a multicarrier phase-based ranging system to any arbitrary value and thus compromise its security. Specifically, we successfully reduced the estimated range to less than $3\unit{m}$ even though the devices were more than 50 m apart. We discussed possible countermeasures that can make it more costly and difficult for an attacker. However, these countermeasures increase the system complexity, do not fully secure against distance decreasing attacks and can be easily circumvented by strong attackers.

\bibliographystyle{abbrv}
\bibliography{phase_ranging}

\begin{thebibliography}{10}

\bibitem{rfranging}
Rf ranging.
\newblock \url{http://www.rfranging.com/}.
\newblock Accessed: 2016-04-09.

\bibitem{USRPN210}
Usrp n210.
\newblock \url{https://www.ettus.com/}.
\newblock Accessed: 2016-04-09.

\bibitem{iso14443}
{ISO/IEC 14443: Identification cards -- Contactless integrated circuit cards --
  Proximity cards -- Part 2: Radio frequency power and signal interface}, 2010.

\bibitem{abrudan2013time}
T.~E. Abrudan, A.~Haghparast, and V.~Koivunen.
\newblock Time synchronization and ranging in ofdm systems using time-reversal.
\newblock {\em IEEE Transactions on Instrumentation and Measurement},
  62(12):3276--3290, 2013.

\bibitem{zigbee}
Z.~Alliance.
\newblock Zigbee.
\newblock Accessed: 2016-04-09.

\bibitem{atmel}
Atmel.
\newblock Atmel avr2152: Rtb evaluation application software user's guide.
\newblock 2013.

\bibitem{BahlMar00}
P.~Bahl and V.~N. Padmanabhan.
\newblock {RADAR: an in-building RF-based user location and tracking system}.
\newblock 2000.

\bibitem{bensky2007}
A.~Bensky.
\newblock {\em Wireless positioning technologies and applications}.
\newblock Artech House, 2007.

\bibitem{BrandsMay93}
S.~Brands and D.~Chaum.
\newblock {Distance-bounding protocols}.
\newblock In {\em Workshop on the theory and application of cryptographic
  techniques on Advances in cryptology}, EUROCRYPT '93, 1993.

\bibitem{clulow2006so}
J.~Clulow, G.~P. Hancke, M.~G. Kuhn, and T.~Moore.
\newblock So near and yet so far: Distance-bounding attacks in wireless
  networks.
\newblock In {\em Security and Privacy in Ad-hoc and Sensor Networks}, pages
  83--97. Springer, 2006.

\bibitem{datasheetsminicircuits}
M.~P.~A. Datasheets.
\newblock Minicircuits products.

\bibitem{exel2013carrier}
R.~Exel.
\newblock Carrier-based ranging in ieee 802.11 wireless local area networks.
\newblock In {\em 2013 IEEE Wireless Communications and Networking Conference
  (WCNC)}, pages 1073--1078. IEEE, 2013.

\bibitem{farnsworth2001high}
B.~D. Farnsworth and D.~W. Taylor.
\newblock High precision narrow-band rf ranging.
\newblock In {\em Proceedings of the 2010 International Technical Meeting of
  The Institute of Navigation}, pages 161--166, 2001.

\bibitem{FluryMar10}
M.~Flury, M.~Poturalski, P.~Papadimitratos, J.-P. Hubaux, and J.-Y.~L. Boudec.
\newblock {Effectiveness of Distance-Decreasing Attacks Against Impulse Radio
  Ranging}.
\newblock 2010.

\bibitem{FrancillonFeb11}
A.~Francillon, B.~Danev, and S.~Capkun.
\newblock {Relay Attacks on Passive Keyless Entry and Start Systems in Modern
  Cars}.
\newblock 2011.

\bibitem{francis2010practical}
L.~Francis, G.~Hancke, K.~Mayes, and K.~Markantonakis.
\newblock Practical nfc peer-to-peer relay attack using mobile phones.
\newblock In {\em Radio Frequency Identification: Security and Privacy Issues}.
  2010.

\bibitem{GuptaMar06}
S.~K.~S. Gupta, T.~Mukherjee, K.~Venkatasubramanian, and T.~B. Taylor.
\newblock {Proximity Based Access Control in Smart-Emergency Departments}.
\newblock 2006.

\bibitem{halperin2011tool}
D.~Halperin, W.~Hu, A.~Sheth, and D.~Wetherall.
\newblock Tool release: gathering 802.11 n traces with channel state
  information.
\newblock {\em ACM SIGCOMM Computer Communication Review}, 41(1):53--53, 2011.

\bibitem{HanckeApr08}
G.~P. Hancke and M.~G. Kuhn.
\newblock {Attacks on time-of-flight Distance Bounding Channels}.
\newblock 2008.

\bibitem{HazasFeb04}
M.~Hazas, J.~Scott, and J.~Krumm.
\newblock {Location-aware computing comes of age}.
\newblock 2004.

\bibitem{6LoWPAN}
L.~IETF.
\newblock Ipv6 over low power wpan.
\newblock Accessed: 2016-04-09.

\bibitem{IEEE802154a07}
The Institute of Electrical and Electronic Engineers.
\newblock {\em {IEEE 802.15.4a-2007 Wireless Medium Access Control (MAC) and
  Physical Layer (PHY) Specifications for Low-Rate Wireless Personal Area
  Networks (WPANs)}}, 2007.

\bibitem{LiuNov07}
H.~Liu, H.~Darabi, P.~Banerjee, and J.~Liu.
\newblock {Survey of Wireless Indoor Positioning Techniques and Systems}.
\newblock 2007.

\bibitem{miesen2012phase}
R.~Miesen, F.~Kirsch, P.~Groeschel, and M.~Vossiek.
\newblock Phase based multi carrier ranging for uhf rfid.
\newblock In {\em Wireless Information Technology and Systems (ICWITS), 2012
  IEEE International Conference on}, 2012.

\bibitem{miesen2013360}
R.~Miesen, A.~Parr, J.~Schleu, and M.~Vossiek.
\newblock 360 degree carrier phase measurement for uhf rfid local positioning.
\newblock In {\em RFID-Technologies and Applications (RFID-TA), 2013 IEEE
  International Conference on}, pages 1--6. IEEE, 2013.

\bibitem{moon2000all}
Y.~Moon, J.~Choi, K.~Lee, D.-K. Jeong, and M.-K. Kim.
\newblock An all-analog multiphase delay-locked loop using a replica delay line
  for wide-range operation and low-jitter performance.
\newblock {\em Solid-State Circuits, IEEE Journal of}, 2000.

\bibitem{PoturalskiSep10}
M.~Poturalski, M.~Flury, P.~Papadimitratos, J.-P. Hubaux, and J.-Y.~L. Boudec.
\newblock {The Cicada Attack: Degradation and Denial of Service in IR Ranging}.
\newblock 2010.

\bibitem{PoturalskiApr11}
M.~Poturalski, M.~Flury, P.~Papadimitratos, J.-P. Hubaux, and J.-Y.~L. Boudec.
\newblock {Distance Bounding with IEEE 802.15.4a: Attacks and Countermeasures}.
\newblock 2011.

\bibitem{ranganathan2015proximity}
A.~Ranganathan, B.~Danev, and S.~Capkun.
\newblock Proximity verification for contactless access control and
  authentication systems.
\newblock In {\em Proceedings of the 31st Annual Computer Security Applications
  Conference}, pages 271--280. ACM, 2015.

\bibitem{RanganathanApr12}
A.~Ranganathan, B.~Danev, A.~Francillon, and S.~Capkun.
\newblock Physical-layer attacks on chirp-based ranging systems.
\newblock In {\em Proceedings of the fifth ACM conference on Security and
  Privacy in Wireless and Mobile Networks}, WISEC '12, 2012.

\bibitem{ranganathan2012design}
A.~Ranganathan, N.~O. Tippenhauer, B.~{\v{S}}kori{\'c}, D.~Singel{\'e}e, and
  S.~{\v{C}}apkun.
\newblock Design and implementation of a terrorist fraud resilient distance
  bounding system.
\newblock In {\em European Symposium on Research in Computer Security}, pages
  415--432. Springer, 2012.

\bibitem{rapinski2015zigbee}
J.~Rapinski and M.~Smieja.
\newblock Zigbee ranging using phase shift measurements.
\newblock {\em Journal of Navigation}, 2015.

\bibitem{rasmussen2010realization}
K.~B. Rasmussen and S.~Capkun.
\newblock Realization of rf distance bounding.
\newblock In {\em USENIX Security Symposium}, pages 389--402, 2010.

\bibitem{RasmussenNov09}
K.~B. Rasmussen, C.~Castelluccia, T.~S. Heydt-Benjamin, and S.~Capkun.
\newblock {Proximity-based Access Control for Implantable Medical Devices}.
\newblock 2009.

\bibitem{RolandSep12}
M.~Roland.
\newblock Applying recent secure element relay attack scenarios to the real
  world: Google wallet relay attack.
\newblock {\em Computing Research Repository}, 2012.

\bibitem{SahinogluDec06}
Z.~Sahinoglu and S.~Gezici.
\newblock {Ranging in the IEEE 802.15.4a Standard}.
\newblock 2006.

\bibitem{salido2013multipath}
D.~Salido-Monz{\'u}, E.~Martin-Gorostiza, J.~Lazaro-Galilea, F.~Domingo-Perez,
  and A.~Wieser.
\newblock Multipath mitigation for a phase-based infrared ranging system
  applied to indoor positioning.
\newblock In {\em Indoor Positioning and Indoor Navigation (IPIN), 2013
  International Conference on}, pages 1--10. IEEE, 2013.

\bibitem{schiller2004location}
J.~Schiller and A.~Voisard.
\newblock {\em Location-based services}.
\newblock Elsevier, 2004.

\bibitem{SedighpourNov05}
S.~Sedighpour, S.~Capkun, S.~Ganeriwal, and M.~B. Srivastava.
\newblock Distance enlargement and reduction attacks on ultrasound ranging.
\newblock 2005.

\bibitem{SpringerApr01}
A.~Springer, W.~Gugler, M.~Huemer, R.~Koller, and R.~Weigel.
\newblock A wireless spread-spectrum communication system using saw chirped
  delay lines.
\newblock 2001.

\bibitem{ubisense10}
Ubisense Technologies.
\newblock {\em {Ubisense Real-time Location Systems (RTLS)}}, 2010.

\bibitem{vasisht2016decimeter}
D.~Vasisht, S.~Kumar, and D.~Katabi.
\newblock Decimeter-level localization with a single wifi access point.
\newblock In {\em 13th USENIX Symposium on Networked Systems Design and
  Implementation (NSDI 16)}, pages 165--178, 2016.

\bibitem{XiangSep04}
Z.~Xiang, S.~Song, J.~Chen, H.~Wang, J.~Huang, and X.~Gao.
\newblock {A wireless LAN-based indoor positioning technology}.
\newblock {\em IBM Journal of Research and Development}, 2004.

\bibitem{xiong2015tonetrack}
J.~Xiong, K.~Sundaresan, and K.~Jamieson.
\newblock Tonetrack: Leveraging frequency-agile radios for time-based indoor
  wireless localization.
\newblock In {\em Proceedings of the 21st Annual International Conference on
  Mobile Computing and Networking}, pages 537--549. ACM, 2015.

\bibitem{zebra10}
Zebra Technologies.
\newblock {\em {Sapphire Dart Ultra-Wideband (UWB) Real Time Locating System}},
  2010.

\bibitem{zhang2014unambiguous}
Y.~Zhang, W.~Qi, and S.~Zhang.
\newblock The unambiguous distance in a phase-based ranging system with hopping
  frequencies.
\newblock {\em arXiv preprint arXiv:1403.1923}, 2014.

\end{thebibliography}
\end{document}